\documentclass{JHEP3}

\usepackage{epsfig}
\usepackage{amsbsy}
\usepackage{varioref}
\usepackage{pifont}

\def\sla#1{\ifmmode%
\setbox0=\hbox{$#1$}%
\setbox1=\hbox to\wd0{\hss$/$\hss}\else%
\setbox0=\hbox{#1}%
\setbox1=\hbox to\wd0{\hss/\hss}\fi%
#1\hskip-\wd0\box1 }

\title{Effects of Fermion Localization in Higgsless Theories and Electroweak Constraints}

\author{Roshan Foadi, Shrihari Gopalakrishna, Carl Schmidt\\
Department of Physics and Astronomy, Michigan State University\\
East Lansing, MI 48824, USA\\
	E-mail: \email{foadi@pa.msu.edu}, 
\email{shri@pa.msu.edu},
\email{schmidt@pa.msu.edu}}

\abstract{Extra-dimensional Higgsless models with electroweak symmetry breaking through boundary conditions generically have difficulties with electroweak precision constraints,
when the fermions are localized to the ``branes'' in the fifth dimension.
In this paper we show that these constraints can be relaxed by allowing the light fermions to
have a finite extent into the bulk of the fifth dimension.   The $T$ and $U$ electroweak 
parameters can be naturally suppressed by a custodial symmetry, while the $S$ parameter 
can be made to vanish through a cancellation, if the leakage into the bulk of the light gauge
fields and the light left-handed fermion fields are of the same size.  This cancellation is
possible while allowing realistic values for the first two generations of fermion masses, although
special treatment is probably required for the top quark.  We present this idea here in the context
of a specific continuum theory-space model; however, it can be 
applied to any five-dimensional Higgsless model, either with a flat or a warped background.
}

\preprint{{~MSUHEP--040917}
{~hep-ph/0409266}}

\begin{document}


\section{Introduction}
\label{sec:Intro}

In the Standard Model (SM), the Higgs sector is responsible for electroweak symmetry breaking. 
The exchange of a virtual Higgs boson perturbatively unitizes the longitudinal gauge boson 
scattering amplitude.   Without a physical Higgs boson, the theory would break down around the TeV scale. 
Higgsless theories have been proposed~\cite{Csaki:2003dt}, as alternatives to the SM, 
in which electroweak symmetry breaking 
is due to boundary conditions on gauge fields that propagate in five dimensions - the usual Minkowskian 
four dimensions plus an additional fifth spatial dimension. 
As is the usual practice, we refer to the extra-dimensional interval as the ``bulk'', and its 
four-dimensional endpoints as ``branes''. 
In Higgsless theories, even though a physical scalar Higgs boson is not present in the theory, it has been shown\cite{SekharChivukula:2001hz} that
the onset of unitarily violation can be delayed due to new contributions from the Kaluza-Klein (KK) excitations of the gauge bosons. 
In our previous work~\cite{FGS} we used deconstruction\cite{Arkani-Hamed:2001ca} to obtain
a Higgsless theory-space model with a $U(1)\times [SU(2)]^N\times SU(2)_{N+1}$ gauge structure.  We found that perturbative unitarily violation could be delayed satisfactorily if the
heavy vector boson states come in below about the TeV scale. 
The continuum limit of this theory-space model was a five-dimensional $SU(2)$ gauge theory
with boundary conditions that break the theory to $U(1)$ on one of the branes and with gauge kinetic terms localized on both branes.

The issue of whether Higgsless theories are compatible with precision electroweak constraints is
being actively investigated. 
In ref.~\cite{Barbieri:2003pr} it was shown that Higgsless theories 
have trouble satisfying precision electroweak constraints, even if brane-localized gauge 
kinetic terms are included.  In our previous work~\cite{FGS}, we showed that in our model, with 
standard model fermions confined to the branes, the contributions to electroweak observables
could be described in terms of the oblique
$S,~T~{\rm and}~U$ parameters~\cite{PT}.   We found that owing to an approximate custodial symmetry, $T$ (and $U$) was compatible with data, 
but $S$ was in violation if the  KK states had masses low enough to satisfy perturbative unitarity.
Possibilities for reducing $S$ in Higgsless theories have been found~\cite{Csaki:2003zu}, but only at the expense of producing a negative value of $T$.
In Ref.~\cite{Chivukula:2004pk} it is claimed that it is not possible to set both the $S$ and $T$ parameters simultaneously to zero, even if the bulk gauge coupling is made position-dependent.  

It is important to note that all of these conclusions about electroweak constraints apply specifically
to Higgsless theories with light fermions bound to the branes.
In this article we shall explore how these conclusions change when the light fermions are allowed to
have some extension into the bulk\footnote{The idea of fermion de-localization
as a potential mechanism to ease constraints from electroweak precision measurements
was mentioned, but not pursued, in Ref.~\cite{Barbieri:2003pr}.}.
We shall use the continuum theory from Ref.~\cite{FGS} as our model, although the basic results should be applicable to any Higgsless theory.
In Section~\ref{sec:model1} we begin by describing the gauge sector, along with a recapitulation of the results from 
Ref.~\cite{FGS} with brane-localized fermions. We then extend this theory 
to incorporate fermions with some finite extension into the bulk.
In Section~\ref{sec:constraints} we show that in this Higgsless model, which contains 
bulk fermions as well as fermion brane kinetic terms, it will be possible to make all of the $S,~T~{\rm and}~U$ parameters small enough to agree with the data. This will be the main result of this article.  Finally, in Section~\ref{sec:conclusions}, we will offer our conclusions and comment
on some remaining issues to be tackled.

\section{Higgsless Theory with Fermions}
\label{sec:model1}

\subsection{Gauge Sector}
\label{subsec:gauge}

As our toy model, we will consider the continuum limit of the theory of ref.~\cite{FGS}, which is 
arguably one of the simplest models of Higgsless electroweak symmetry breaking.
This model is an $SU(2)$ gauge theory, defined on a fifth-dimensional line segment, 
$0\leq y\leq\pi R$, where the boundary conditions break the gauge symmetry down to $U(1)$
at one end of the interval. The five dimensional action is\footnote{Note that we have taken
 $y\leftrightarrow \pi R -y$ with respect to the action in ref.~\cite{FGS}. We have also scaled out a
 factor $\pi R$ in the first term in order to make $\hat{g}_5$ dimensionless.}
\begin{eqnarray}
{\cal S}&=& \int_0^{\pi R}dy\int d^4x
\left[
-{1\over4(\pi R)\hat{g}_5^2}W^{a\,MN}W^a_{MN}
-\delta(y){1\over4g^2}W^{a\,\mu\nu}W^a_{\mu\nu}\right.\nonumber\\
&&\qquad\qquad\qquad\quad\left.-\delta(\pi R-y)
{1\over4g^{\prime2}}W^{3\,\mu\nu}W^3_{\mu\nu}\right]\ ,
\label{eq:5daction}
\end{eqnarray}
where, in this equation, the indices $M,N$ run over the 5 dimensions,
and we impose the Dirichlet Boundary condition, $W^a_\mu=0$, 
at $y=\pi R$ for $a\ne3$.  The boundary kinetic energy term at $y=0$ is defined by
interpreting the $\delta$-function as $\delta(y-\epsilon)$ with 
$\epsilon\rightarrow0^+$ and the fields having Neumann boundary conditions,
 $dW^a_\mu/dy=0$, at $y=0$.  The $\delta$-function 
and the field $W^3_\mu$ at $y=\pi R$ should be interpreted similarly.
Note that in the limit of small $g$ and $g^\prime$ the theory looks like an $SU(2)$ gauge theory
and  a $U(1)$ gauge theory, living at the left and right ends of the fifth-dimensional 
interval, respectively.  It is the bulk fields which connect the $SU(2)$ and the $U(1)$ theories,
and transmit the breaking of the gauge groups down to a single $U(1)_{EM}$.

The five-dimensional gauge fields can be expanded in a tower of four-dimensional Kaluza-Klein
(KK) states:
\begin{eqnarray}
W^{\pm\mu}(x,y)&=& \sum_{n=0}^\infty f_n(y)W_n^{\pm\mu}(x)\nonumber\\
W^{3\mu}(x,y)&=& eA^\mu(x)+\sum_{n=0}^\infty g_n(y)Z_n^{\mu}(x)\ ,
\label{eq:gaugeKK}
\end{eqnarray}
where $W_n^{\pm\mu}$ has mass $m_{W_n}$, and
$Z_n^{\mu}$ has mass $m_{Z_n}$. 
The lowest states of the tower, 
$W_0^{\pm\mu}$ and $Z_0^\mu$, are identified as the standard model $W^\pm$
and $Z$ bosons, respectively.   
Solving for the masses perturbatively in $\lambda\equiv g/\hat{g}_5$ and $\lambda'\equiv g'/\hat{g}_5$ we obtain
\begin{eqnarray}
m_W^2\equiv m_{W_0}^2 & = & {\lambda^2\over (\pi R)^2}\left(1-{\lambda^2\over 3}
+{\cal O}(\lambda^4)\right) \nonumber \\
m_Z^2\equiv m_{Z_0}^2 & = & {\lambda^2+\lambda'^2\over (\pi R)^2}\left(1-{\lambda^2+\lambda'^2\over 3}
+{\lambda^2\lambda'^2\over\lambda^2+\lambda'^2}+{\cal O}(\lambda^4)\right) 
\label{eq:smmass}
\end{eqnarray}
for the standard model gauge bosons, and
\begin{eqnarray}
m_{W_n}^2&\approx&m_{Z_n}^2 \ = \ \left({n\over R}\right)^2\left(1
+{\cal O}(\lambda^2)\right) 
\label{eq:gaugemass}
\end{eqnarray}
for the heavy gauge bosons.

\subsection{Fermion Sector Model I:  Brane-localized Fermions}
\label{subsec:fermion1}

For clarity we first consider the fermion sector of ref.~\cite{FGS},
where the fermions are restricted to the branes.  The continuum limit of the fermion action can be written
\begin{eqnarray}
{\cal S}^{(I)}&=& \int_0^{\pi R}dy\int d^4x
\left[
\delta(y)i\bar{\psi}_L\sla{D}\psi_L+\delta(\pi R-y)i\bar{\psi}_R\sla{D}\psi_R
\right]\ ,
\label{eq:fermionaction1}
\end{eqnarray}
where the covariant derivatives on the left- and right-handed fields are
\begin{eqnarray}
\sla{D}\psi_L&=& \left(\sla{\partial}-iT^a\sla{W}^a(y)-iY_L\sla{W}^3(\pi R)\right)\psi_L\nonumber\\
\sla{D}\psi_R&=& \left(\sla{\partial}-iY_R\sla{W}^3(y)\right)\psi_R\ .
\label{eq:derivs1}
\end{eqnarray}
Note that the left-handed field $\psi_L$ lives at $y=0$ 
but couples also to the gauge field $W^3$ at $y=\pi R$.  This non-locality may seem unnatural
from the standpoint of an extra-dimensional theory; however, it is perfectly well-defined if we
consider this from the standpoint of a continuum theory-space model.   In the theory-space
interpretation $W^a(0)$ and $W^3(\pi R)$ are just the gauge fields for independent $SU(2)$
and $U(1)$ gauge groups, and $y$ is a (continuous) label for the independent gauge groups.

Unfortunately, this mode of incorporating fermions has some unpleasant features. 
In order to give mass to the fermions requires introducing a nonlocal
mass term involving a Wilson line running between the two branes.  But the
most damaging feature of this fermion action is that it produces electroweak radiative 
corrections that are too large, invalidating the theory.  Therefore, we now consider an
alternative fermion action.

\subsection{Fermion Sector Model II: Fermions with Finite Extension into the Bulk}
\label{sec:fermions2}

Drawing on the analogy of the gauge action (\ref{eq:5daction}), which has $SU(2)$ and $U(1)$
kinetic terms peaked at the two ends of the interval and connected through the bulk kinetic
term, we consider a theory with left-handed and right-handed fermion
kinetic terms peaked at the two ends of the
interval and connected through a bulk fermion kinetic term\footnote{Note that we have scaled out a factor
$\pi R$ in the bulk integral, in order for the parameters $t_L$, $t_{\nu_R}$, and $t_{e_R}$to be 
dimensionless.}:
\begin{eqnarray}
{\cal S}^{(II)} = \int_0^{\pi R}dy && \int d^4x
\left[{1\over\pi R}\left({i\over2}\bar{\psi}\Gamma^M D_M\psi + h.c.
-M\bar{\psi}\psi\right)\right.\nonumber\\
&&\left.
+\delta(y){1\over t_L^2}i\bar{\psi}_L\sla{D}\psi_L
+\delta(\pi R-y)\left({1\over t_{\nu_R}^2}i\bar{\nu}_R\sla{D}\nu_R
+{1\over t_{e_R}^2}i\bar{e}_R\sla{D}e_R\right)
\right]\ ,
\label{eq:fermionaction2}
\end{eqnarray}
The five-dimensional Dirac matrices are defined in terms of the four-dimensional ones by
$\Gamma^M = (\gamma^\mu,-i\gamma^5)$.  The five-dimensional fermion is equivalent to
a four-dimensional Dirac fermion, $\psi=(\psi_L,\psi_R)$, where $\psi_L$ and $\psi_R$ are $SU(2)$
doublets, $\psi_L=(\nu_L,e_L)$ and $\psi_R=(\nu_R,e_R)$. 
The boundary kinetic energy term at $y=0$ is defined by
interpreting the $\delta$-function as $\delta(y-\epsilon)$ for $\epsilon\rightarrow0^+$
with the boundary condition
$\psi_R=0$ at $y=0$.  Similarly, the boundary  term at $y=\pi R$ is defined by
interpreting the $\delta$-function as $\delta(\pi R-y+\epsilon)$ with the boundary condition
$\psi_L=0$ at $y=\pi R$.  The general treatment of possible fermion boundary conditions
can be found in Ref.~\cite{Csaki:2003sh}.

The covariant derivative in Eq.~(\ref{eq:fermionaction2}) is
\begin{equation}
D_M\psi = \left(\partial_M - iT^aW_M^a(y)-iY_LW_{M}^3(\pi R)\right)\psi,
\label{eq:bulkderiv}
\end{equation}
where $Y_L$ is the $\psi_L$ hypercharge. At the interval
boundaries the four-dimensional part of the covariant derivative (\ref{eq:bulkderiv}) becomes:
\begin{eqnarray}
(\sla{D}\psi_L)_{y=0} & = & \left(\sla{\partial}-iT^a\sla{W}^a(0)-iY_L\sla{W}^3(\pi R)\right)\psi_L \nonumber \\
(\sla{D}\psi_R)_{y=\pi R} & = & \left(\sla{\partial}-iT^3\sla{W}^3(\pi R)-iY_L\sla{W}^3(\pi R)\right)\psi_R \nonumber \\
& = & \left(\sla{\partial}-iY_R\sla{W}^3(\pi R)\right)\psi_R \ ,
\label{eq:derivends}
\end{eqnarray}
where the $\psi_R$ hypercharge, $Y_R$, is related to $Y_L$ by $Y_R=T^3+Y_L$, as in the SM. Note that 
$Y_R$ is a 2$\times$2 diagonal matrix, with the $\nu_R$ hypercharge on the upper left, 
and the $e_R$ hypercharge on the
lower right. 
Therefore, at $y=\pi R$ the covariant derivative term, $\bar{\psi}_R\sla{D}\psi_R$, 
splits into two 
separately gauge invariant terms, $\bar{\nu}_R\sla{D}\nu_R$ and $\bar{e}_R\sla{D}e_R$, as in 
Eq.(\ref{eq:fermionaction2}). Note also that in the limit of small $t_L$, $t_{\nu_R}$, and $t_{e_R}$ 
 the 
action ${\cal S}^{(II)}$ describes massless left-handed fermions gauged under an
 $SU(2)\times U(1)$ group living on the left end of the fifth-dimensional interval, and 
massless right-handed fermions gauged under a $U(1)$ living on the right end of the interval, 
exactly as in model I.  
It is the presence of the bulk fields which allow these light states to communicate with each
other, supplying the analog of the Yukawa coupling of the SM, and giving mass to the fermions.

Letting $\chi$ denote either the five-dimensional electron or neutrino field, we can
 expand in a tower of four-dimensional KK states:
\begin{eqnarray}
\chi_L(x,y)&=& \sum_{n=0}^\infty \alpha_n(y)\chi_{nL}(x)\nonumber\\
\chi_R(x,y)&=& \sum_{n=0}^\infty \beta_n(y)\chi_{nR}(x) \ .
\label{eq:fermionKK}
\end{eqnarray}
The four dimensional fields $\chi_{nL}$ and $\chi_{nR}$ are 
the left-handed and right-handed components, respectively, of a mass-$m_n$ Dirac fermion, 
$(\chi_{nL},\chi_{nR})$.   
The heavy fermions, labeled by $n>0$ are just standard bulk fermions with masses
determined by the bulk mass and the boundary conditions, and only slightly perturbed by
the brane terms.  The light fermions, however, are dominated by the brane terms, with only 
small extension into the bulk set by $t_L$ and $t_{\chi_R}$, which allows a light mass to 
exist.
Solving perturbatively in $t_L$ and $t_{\chi_R}$ we find a light Dirac state with mass
\begin{equation}
m_{\chi_0} = {t_L t_{\chi_R}e^{-(M\pi R)}\over\pi R}\left(1+ {\cal O}(t^2)
\right) \ .
\label{eq:mass2}
\end{equation}

The fermion sector of model II can accomodate both leptons and quarks, as well as multiple generations and generational mixing.
In this work we shall assume that $t_L$ and the bulk mass $M$ are universal for all fermions, 
so that the masses and mixing are determined by the $t_{\chi_R}$.  Assuming $t_L$ to
be of the same size as $\lambda\equiv g/\hat{g}_5$ and $M$ not too large, this implies that
the $t_{\chi_R}$ are very small, except for the third generation.  
For example, with $R^{-1}\sim1$ TeV and $t_L\sim \lambda\sim10^{-1}$, we find that
$t_{\chi_R}$ ranges from $10^{-11}$ for the lightest neutrino to $10^{-2}$ for the charm quark.
We shall postpone the 
discussion of the details of the fermion masses and mixings, as well as problems associated
 with the third generation, to a followup paper~\cite{followup}.  For the remainder of this article
we assume that $t_{\chi_R}$ is neglible and examine the effects of a universal and small, but
non-neglible, value of $t_L$.

\section{Electroweak Constraints}
\label{sec:constraints}

In Ref.~\cite{FGS} we showed that the electroweak precision constraints in model I 
with brane-localized fermions could be parametrized to order $\lambda^2$ fully in terms of the oblique parameters $S$, $T$, and $U$~\cite{PT}.
This was possible because the couplings between the light fermions and the new heavy
vector bosons are suppressed by a factor $\lambda$ relative to the couplings to the standard
model $W$ and $Z$.  As a result, the contributions of the heavy vector bosons to 
four-fermion operators at zero-momentum transfer are suppressed by a factor of $\lambda^4$
relative to the standard model $W$ and $Z$ contributions, and can be neglected.  
(There is a relative factor of
$\lambda^2$ from the couplings and an additional factor of $\lambda^2$ due to the
the heavy vector boson masses in the boson propagator.)  To order $\lambda^2$, 
we only need consider the couplings to the standard model $W$ and $Z$ for electroweak 
precision measurements.  Any additional universal parameters, such as those considered in Ref.~\cite{Barbieri:2004qk}
can be neglected.

This result also applies to model II with light fermion extension into the bulk, if we make the two simple 
assumptions that we introduced earlier.
Firstly, if we assume that $t_L\approx\lambda$, then the coupling between light fermions and
the heavy vector bosons remains suppressed by a factor of $\lambda$, although the coefficient
multiplying this factor is changed.  (Effectively, the enhanced overlap of the bulk fermion wavefunctions with the heavy vector boson wavefunctions is compensated by the small probability for the light fermions to leak into the bulk, producing a change in this coefficient of order $t_L^2/\lambda^2\sim1$.)  Secondly, if we assume that $t_{\chi_R}$ is negligible, then
there are no additional right-handed charged or neutral currents beyond those that occur in the
standard model.  The couplings of the light fermions and the standard model $W$ and $Z$ can be described by the following interaction lagrangians,
\begin{eqnarray}
{\cal L}_{CC}&=& {g^{CC}\over\sqrt{2}}W^{+\mu}\,\bar{\nu}\gamma_\mu P_Le\ +\ {\rm h.c.}\nonumber\\
{\cal L}_{NC}&=& Z^{\mu}\,\bar{\psi}
\left[g^{NC}_{3}T^3\gamma_\mu P_L
+g^{NC}_{Q}Q\gamma_\mu 
\right]\psi\ +\ {\rm h.c.}\ .
\label{eq:CCLag}
\end{eqnarray}

With these assumptions, we can now parametrize the influence of the new physics on the couplings in terms of $S$, $T$, and $U$.  We shall take $\alpha$, $m_W$, and $m_Z$ as the
fundamental input observables, since their relation to the parameters in the lagrangian
is independent of how the fermions are incorporated into the theory.  This is trivially seen to be
true for $m_W$ and $m_Z$, while the flatness of the photon wave function imposes that
\begin{eqnarray}
e^2&=& \left({1\over \hat{g}_5^2}+{1\over g^2}+{1\over g^{\prime2}}\right)^{-1}\ ,
\label{eq:em1}
\end{eqnarray}
independent of fermion model.  
 Following ref.~\cite{burgess}, we find the deviations in the relations between the universal 
$W$ and $Z$ couplings to be
\begin{eqnarray}
g^{CC}&=& {e\over s}\left[1+{\alpha S\over4s^2}-
{c^2\alpha T\over2s^2}-{(c^2-s^2)\alpha U\over8s^4}\right]\nonumber\\
g^{NC}_{3}&=& {e\over sc}\left[1+{\alpha S\over4s^2}-
{(c^2-s^2)\alpha T\over2s^2}-{(c^2-s^2)\alpha U\over8s^4}\right]
\nonumber\\
g^{NC}_{Q}&=& -{es\over c}\left[1+{\alpha T\over2s^2}+{\alpha U\over8s^4}\right]
\ .
\label{eq:STUdef}
\end{eqnarray}
In these expressions we have defined $c\equiv m_W/m_Z$ and $s\equiv(1-c^2)^{1/2}$.
Note that our choice of definition for $\sin^2{\theta_W}$ is different from that used in Ref.~\cite{burgess}.

In model I with brane-localized fermions the couplings are determined by the values of the light boson wavefunctions at the boundaries
\begin{eqnarray}
g^{CC(I)}&=& f_0(0)\nonumber\\
g^{NC(I)}_{3}&=&g_0(0)-g_0(\pi R)\nonumber\\
g^{NC(I)}_{Q}&=& g_0(\pi R)\ .
\label{eq:couplingsWF1}
\end{eqnarray}
Solving perturbatively for these normalized wave functions, we find
\begin{eqnarray}
g^{CC(I)}&=& {e\over s}\left[1+\lambda^2/6 +{\cal O}(\lambda^4)\right]\nonumber\\
g^{NC(I)}_{3}&=& {e\over sc}\left[1+\lambda^2/6 +{\cal O}(\lambda^4)\right]\nonumber\\
g^{NC(I)}_{Q}&=& -{es\over c}\left[1+{\cal O}(\lambda^4)\right]
\ .
\label{eq:couplings1}
\end{eqnarray}
Thus, we obtain for this theory
\begin{eqnarray}
\alpha S&=& 2s^2\lambda^2/3\nonumber\\
\alpha T&=&0\nonumber\\
\alpha U&=&0\ ,
\label{eq:STU1}
\end{eqnarray}
as previously found in Ref.~\cite{FGS}.

We now consider the couplings in fermion model II with finite extension into the bulk.  
They can be expressed as
\begin{eqnarray}
g^{CC(II)}&=&g^{CC(I)}  \int_0^{\pi R} dy\left[{1\over\pi R}+{1\over t_L^2}\delta(y)\right]
\left({f_0(y)\over f_0(0)}\right)\alpha(y)^2\nonumber\\
g^{NC(II)}_{3}&=&g^{NC(I)}_{3}\int_0^{\pi R} dy\left[{1\over\pi R}+{1\over t_L^2}\delta(y)\right]
\left({g_0(y)-g_0(\pi R)\over g_0(0)-g_0(\pi R)}\right)\alpha(y)^2\nonumber\\
g^{NC(II)}_{Q}&=& g^{NC(I)}_{Q}\ ,
\label{eq:couplings2p}
\end{eqnarray}
where we have factored out the values from model I for clarity.  (Note that we have
assumed that all of the left-handed fermion wave functions are identical, which is valid
in the limit of negligible $t_{\chi_R}$.)
The change in the $g^{CC}$ and $g^{NC}_3$ couplings between model I and model II is, in fact, 
a suppression factor. This is seen from the fact that
the fermion wave functions are normalized by
\begin{eqnarray}
1\ =\ \int_0^{\pi R} dy\left[{1\over\pi R}+{1\over t_L^2}\delta(y)\right]
\alpha_(y)^2\ ,
\label{eq:fermnorm}
\end{eqnarray}
while the factors in parentheses are positive and less than one:
\begin{equation}
0\le\qquad {f_0(y)\over f_0(0)} \approx {g_0(y)-g_0(\pi R)\over g_0(0)-g_0(\pi R)}
\approx 1-{y\over\pi R}\qquad \le1\ .
\end{equation}
The suppression factors for $g^{CC}$ and $g^{NC}_3$ are identical to leading order in $\lambda^2$.

Evaluating the integrals, we obtain
\begin{eqnarray}
g^{CC(II)}&=&g^{CC(I)}  (1-At_L^2)\nonumber\\
g^{NC(II)}_{3}&=&g^{NC(I)}_{3}(1-At_L^2)\nonumber\\
g^{NC(II)}_{Q}&=& g^{NC(I)}_{Q}\ ,
\label{eq:couplings2}
\end{eqnarray}
where
\begin{eqnarray}
A&=&e^{-\hat{M}}{\sinh\hat{M}\over\hat{M}}
-{1\over2\hat{M}}\left(1-e^{-\hat{M}}{\sinh\hat{M}\over\hat{M}}\right)
\ ,
\label{eq:CCsm}
\end{eqnarray}
and $\hat{M}=M\pi R$ is the scaled bulk mass.  
In the limit $\hat{M}\rightarrow0$ we find $A\rightarrow1/2$ .
We now see that allowing the fermions to extend into the bulk, as in model II, can be used
to cancel the effects of $S$ in electroweak measurements.  Comparing Eq.~(\ref{eq:couplings2})
with Eq.~(\ref{eq:couplings1}), we see that $S$ can effectively be set to zero (while retaining $T=U=0$) by the choice
\begin{equation}
t_L^2\ =\ {\lambda^2\over6A}\ .
\end{equation}

\section{Conclusions}
\label{sec:conclusions}
In models of Higgsless electroweak symmetry breaking
it is straightforward to incorporate a custodial symmetry, which naturally ensures that $T=0$.
However, if the standard model fermions are localized to the branes, the contribution to 
$S$ is typically sizable.  In the analysis of Ref.~\cite{FGS} and reproduced as model I in
 this paper, the contribution to $S$ is proportional to $\lambda^2$, where $\lambda$ is the ratio
of the brane to bulk gauge couplings.   The quantity $\lambda^2$
is fixed to be of order $10^{-2}$ by the fact that the standard model gauge bosons
have the right mass, while the first KK vector boson must have a mass below about 1 TeV in order to preserve unitarity up to some reasonable scale.  

In this paper we have extended the analysis of Ref.~\cite{FGS} to include the effects of
light fermion extension into the bulk with large fermion brane kinetic terms in model II (with coefficient $1/t_L^{2}$), in direct analogy to the gauge sector.  This has no effect
on the custodial symmetry, thus keeping $T=0$, but it does produce a new contribution
to $S$ which is proportional to $t_L^2$ and of opposite sign to the previous contribution
proportional to $\lambda^2$.  Therefore, it is possible to obtain a cancellation between these
contributions.  Of course, to obtain $S=0$ would appear to require fine tuning of two independent
parameters, $\lambda^2$ and $t_L^2$.  At the moment, we do not have a precise mechanism
to produce this tuning naturally; however, we do note that the quantities play suggestively analogous
roles in the two different sectors of the theory.  That is, the quantity $\lambda^2$ is 
a measure of the extent to which the gauge fields leak into the bulk away from the brane at 
$y=0$, while the quantity $t_L^2$ is a measure of the extent to which the left-handed fermions 
leak into the bulk away from the brane at $y=0$.  This offers hope that these parameters might
be correlated and cancel naturally in some future model.  In any event, we offer this as a proof
of principle of a Higgsless model with both $S$ and $T$ set to zero.

In model II the light fermion masses-squared are suppressed with respect
to the natural size of $(\pi R)^{-2}$ by the amount that the left-handed and right-handed
fermions can leak away from the branes at $y=0$ and $y=\pi R$ respectively.  That is,
they are suppressed by the factor $t_L^2t_{\chi_R}^2$.  Given that $t_L$ is constrained
to keep $S$ fixed at zero, the masses are determined by $t_{\chi_R}$. 
We have found that all the SM-fermion masses can be obtained, while keeping $t_L$ universal, except for the case of the top quark. The corresponding values of $t_{\chi_R}$ are negligible in the
calculation of electroweak observables. The details of the mass spectrum, as well as a discussion
of the top-quark problem will be postponed to a followup paper~\cite{followup}.

Finally, we want to stress the generality of this approach. For simplicity, we have considered a model from continuum theory space. However, this picture can be generalized to any 5-dimensional models, with flat or warped backgrounds. For example,
one can apply it to the model of Ref.~\cite{Barbieri:2003pr}, by noting that their model can be
obtained as a generalization of our continuum theory space model.  If we start with our model
and ``fold it in half'' by identifying the points $y$ and $\pi R-y$, we can treat the $SU(2)$
symmetries for  $y<\pi R/2$ and $y>\pi R/2$ as independent gauge groups, $SU(2)_L$ and $SU(2)_R$ (connected by an appropriate boundary condition at the reflection point $y=\pi R/2$).  The $U(1)$ symmetry at $y=\pi R$ in our model can then be extended to range over 
$0\le y\le\pi R/2$
to obtain the model of Ref.~\cite{Barbieri:2003pr}.  This same ``folding'' procedure can also
be applied to the fermions in our model II, producing two sets of bulk fermions, both with
boundary kinetic terms at $y=0$, and connected by an appropriate boundary condition
at $y=\pi R/2$.  Given the generality of this approach to satisfying the electroweak precision
constraints, further investigations in this direction are worthwhile, with particular attention to the
constraints imposed by top-quark phenomenology.

While we were completing this manuscript, a paper~\cite{Cacciapaglia:2004rb} 
appeared on the ArXiv that discusses localization of
fermions as a way to address electroweak constraints in the context of a Higgsless model in warped space.  The general
conclusions in that paper are similar to ours, although the details are quite different.  They also include some discussion
of flavor issues and difficulties with the third generation in the warped-space model.

\vskip .2 cm

\section*{Acknowledgments}

This work was supported by the US National
Science Foundation under grants PHY-0070443
and PHY-0244789.   We would like to thank Sekhar Chivukula for useful discussions.


\end{document}